\documentstyle[12pt]{article} 
\input{psfig}
\begin{document}
 
\author{{\bf G. Herrera} 
\thanks{e-mail: gherrera@fis.cinvestav.mx} \\ 
{\small Centro de Investigaci\'{o}n y de Estudios Avanzados} \\ 
{\small Apdo. Postal 14-740, M\'{e}xico 07000, DF, Mexico} \\ 
\\
\and {\bf J. Magnin}
\thanks{e-mail: jmagnin@lafex.cbpf.br} \\ 
{\small Centro Brasileiro de Pesquisas F\'{\i}sicas} \\ 
{\small Rua Dr. Xavier Sigaud 150, CEP 22290-180, Rio de Janeiro, Brazil}}

\title{$\Lambda _c/\overline{\Lambda}_c$ production asymmetries 
in $pp$ and $\pi^{-}p$ collisions}

\date{}
\maketitle

\begin{tabular}{rrl}
PACS & 13.60.Rj & (Baryon production)\\
     & 13.87.Fh & (Fragmentation into hadrons)\\
     & 14.20.Lq & (Charmed baryons)
\end{tabular}

\begin{abstract}
We study $\Lambda_c/\overline{\Lambda}_c$ production asymmetries 
in $pp$ and $\pi^{-}p$ collisions using a recently proposed 
two component model. The model includes heavy baryon production by 
the usual mechanism of parton fusion and fragmentation plus 
recombination of valence and sea quarks from the beam and 
target hadrons. We compare our results with experimental data
on asymmetries measured recently.
\end{abstract}

\newpage
\section{Introduction}

Recent experimental data on charmed meson hadroproduction shows a 
strong correlation between the flavor content of the incident hadron 
and the produced meson \cite{ramona0,e791dd}. This effect, known as leading 
particle effect, has been observed also in $\Lambda_c$ 
production in $pp$ 
\cite{ramona1,e769} and $\pi^- p$ 
\cite{e791} interactions, in $\Lambda_b$ production in $pp$ 
collisions \cite{ramona2} and $\Xi_c$ production in hyperon-nucleus 
interactions \cite{ramona3}.

The leading particle effect can not be explained within the 
usual framework based in the factorization theorem. In fact, 
although perturbative QCD (pQCD) at Next to Leading Order (NLO) 
predicts a small enhancement of the $\bar{c}$ over the $c$-quark 
cross section \cite{nlo0}, the effect of this asymmetry is very 
small to account for the observed $D^-/D^+$ asymmetry measured 
in $\pi^-$-nucleus interactions \cite{e791dd}. On the other hand, 
the evidence of leading particle effects in $\Lambda_c$ production 
at large $x_F$ clearly shows that non-perturbative mechanisms 
play a fundamental role in hadroproduction.

Leading particle effects in  charmed hadron 
production has been studied from two 
different points of view: the soft scattering - low virtuality 
approach \cite{hwa-prd}, in which charmed meson anti-meson production is 
reviewed within the valon recombination model, and two component models, 
in which charmed hadron production takes place by two different 
mechanisms, namely perturbative QCD (pQCD) 
followed by fragmentation plus coalescence of 
intrinsic charm \cite{ramona4} or recombination of the valence quarks 
with the charmed sea quarks of the initial hadrons liberated in 
the collision \cite{nosotros}. In the two component models, the intrinsic 
charm coalescence or the valence and sea quark recombination gives 
the non-perturbative contribution which takes into account the observed 
flavor correlation.

Leading particle correlations can be quantified by studies of the 
production asymmetries between leading and non-leading particles. The 
production asymmetry is defined by
\begin{equation}
A_{L/NL}(x_F) = \frac{d\sigma_L/dx_F - 
d\sigma_{NL}/dx_F}{d\sigma_L/dx_F + 
d\sigma_{NL}/dx_F}
\label{eq0}
\end{equation}
in which L stands for Leading and NL for Non-Leading particles.
In this paper we give the predictions of the recombination two component 
model for $\Lambda_c/\overline{\Lambda}_c$ production asymmetries 
in $pp$ and $\pi^- p$ interactions. We also compare the predictions of the 
model with recent experimental data in $pp$ \cite{e769} 
and $\pi^- p$ \cite{e791} interactions.

\section{$\Lambda_c$ and $\overline{\Lambda}_c$ production in $pp$ 
and $\pi^- p$ collisions}

In this section we review $\Lambda_c/\overline{\Lambda}_c$ production 
in the recombination two component model \cite{nosotros}. In section 
\ref{qcd} we present the calculation of the parton fusion contribution 
and in section \ref{recom} the recombination picture is studied for both 
$pp$ and $\pi^- p$ interactions.

\subsection{$\Lambda_c$ and $\overline{\Lambda}_c$ production 
{\it via} parton fusion}
\label{qcd} 

The calculation we present here is at Lowest Order (LO) in $\alpha_s$. 
A constant factor $K \sim 2-3$ is included in the parton fusion cross 
section to take into account Next to Leading Order (NLO) 
contributions \cite{nlo}.

\begin{figure}[b] 
\psfig{figure=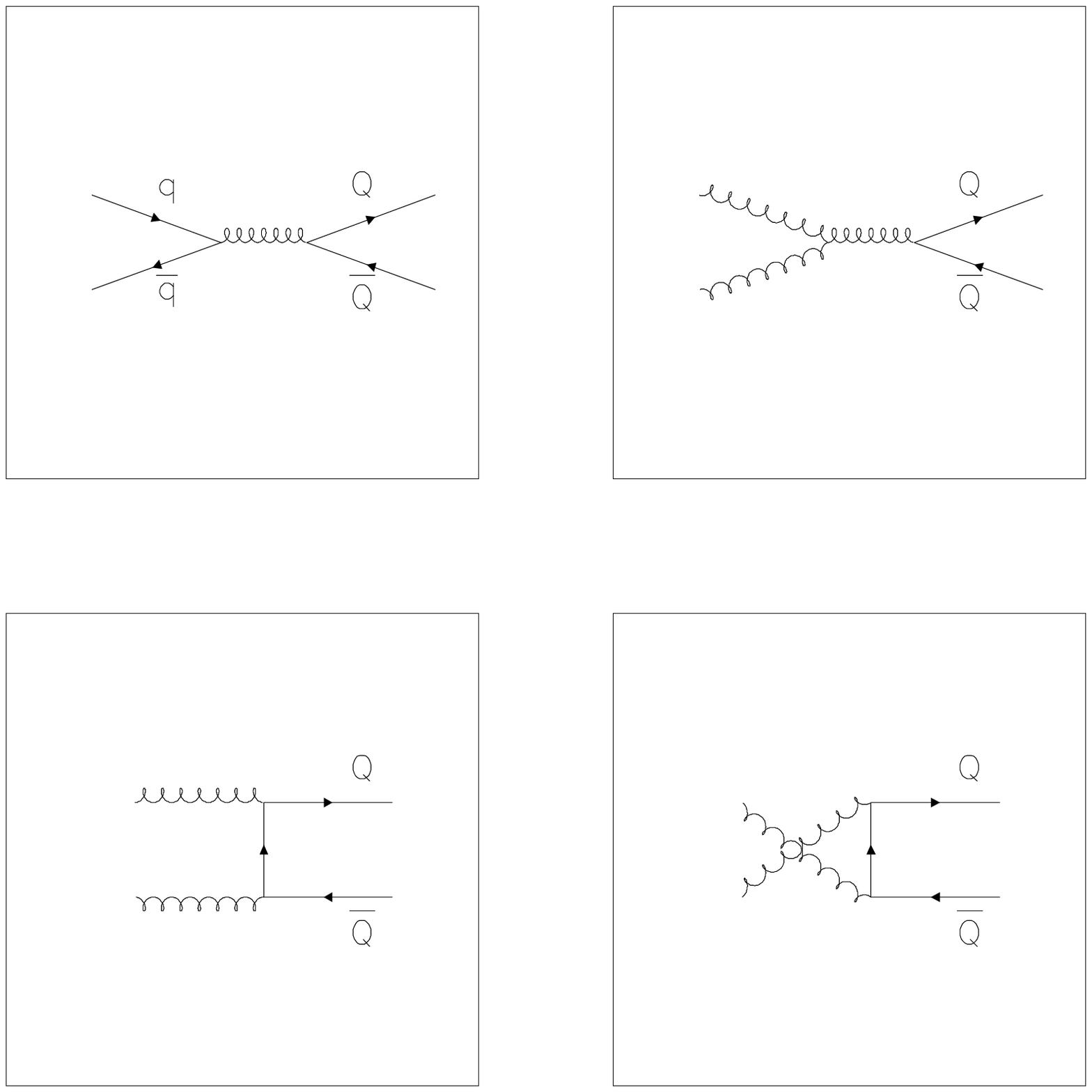,height=6.0in} 
\caption{}
\label{feynman} 
\end{figure}

The inclusive $x_F \left(=2p_L/\sqrt{s}\right)$ distribution for 
$\Lambda_c$ $(\overline{\Lambda}_c)$ production in hadron-hadron 
interactions, assuming factorization, has the form \cite{vbh-npb}
\begin{equation}
\frac{d\sigma^{pf} }{dx_F}=\frac{1}{2} \sqrt{s} \int H_{ab}^{AB}
(x_a,x_b,Q^2)
\frac{1}{E} \frac{D_{\Lambda_c/c} \left( z \right)}{z} dz dp_T^2 dy \: ,
\label{sig-qcd} 
\end{equation}
where
\begin{eqnarray}
H_{ab}^{AB}(x_a,x_b,Q^2)& = & \Sigma_{a,b} \left( q_a(x_a,Q^2)
\bar{q_b}(x_b,Q^2)\right. \nonumber \\
                   &   & + \left. \bar{q_a}(x_a,Q^2) q_b(x_b,Q^2) \right)
\frac{d \hat{\sigma}}{d \hat{t}} \left|_{q\bar{q}} \right. \nonumber \\
                   &   & + g_a(x_a,Q^2) g_b(x_b,Q^2) \frac{d \hat{\sigma}}
{d \hat{t}}\left|_{gg}\right. \;,
\label{int-qcd}
\end{eqnarray}
with
$x_a$ and $x_b$ the parton momentum fractions in the inital hadrons A 
and B, $q(x,Q^2)$ and $g(x,Q^2)$ the quark and gluon distribution 
in the corresponding colliding hadrons, $E$ the energy of
the produced $c \left(\bar c\right)$-quark and $D_{\Lambda_c/c} \left( z
\right)$ the appropriated fragmentation function. In eq. \ref{sig-qcd},
$p_T ^2$ is the squared transverse momentum of the produced $c \left(\bar
c\right)$-quark, $y$ is the rapidity of the $\bar {c}\left(c\right)$ quark
and $z=x_F/x_c$ is the momentum fraction of the charm quark carried by the
$\Lambda _{c}\left(\overline{\Lambda}_c\right)$. The sum in eq. 
\ref{int-qcd} runs over $a,b = u,\bar{u},d,\bar{d},s,\bar{s}$.
$d \hat{\sigma}/d \hat{t} \mid_{q\bar{q}}$ and 
$d\hat{\sigma}/d \hat{t} \mid_{gg}$ are the elementary cross sections for
the hard processes $q\bar{q} \rightarrow c\bar c$ and $gg\rightarrow c\bar
c$ at LO given by
\begin{equation} \frac{d
\hat{\sigma}}{d \hat{t}}\mid_{q \bar {q}} = \frac{\pi \alpha _{s}^{2}
 \left( Q^2 \right)}{9 \hat{m}_{c}^{4}} \;  \frac{cosh \left( \Delta y
\right) + m_{c}^{2}/ \hat{m}_{c}^{2}} {\left[ 1+cosh \left( \Delta y
\right) \right] ^3} \label{q-antiq} 
\end{equation} 
\begin{equation}
\frac{d \hat{\sigma}}{d \hat{t}}\mid_{gg}= \frac{\pi \alpha_{s}^{2} \left(
Q^2 \right)}{96 \hat{m}_{c}^{4}} \;  \frac{8 cosh \left( \Delta y \right)
-1}{\left[ 1+cosh \left( \Delta y \right) \right]^3} \: \left[ cosh \left(
\Delta y \right)+
\frac{2m_c^2}{\hat{m}_c^2}+\frac{2m_c^4}{\hat{m}_c^4}\right], \label{gg}
\end{equation} 
where $\Delta y$ is the rapidity gap between the produced
$c$ and $\bar{c}$ quarks and $\hat{m}_c^2=m_c^2+p_T^2$.  The Feynman
diagrams involved in the calculation of eqs. \ref{q-antiq} and \ref{gg}
are shown in Fig. \ref{feynman}.

For consistency with the LO calculation, we use the
GRV-LO parton distributions for both the proton \cite{grv-p} and the pion
\cite{grv-pi}. The hard momentum scale is fixed at $Q^2 = 2m_c^2$.\\
The fragmentation is modeled by two different functions; the
Peterson fragmentation function extracted from data in
$e^+ e^-$ interactions \cite{peterson} 
\begin{equation} D_{\Lambda_c/c}(z)
= \frac{N}{z \left[ 1 - 1/z - \epsilon_c/(1-z) \right]^2}
 \label{peter}
\end{equation}
with $\epsilon_c = 0.06$ and the normalization defined by 
$\sum _{H} \int D_{H/c}(z) dz = 1$ and the delta fragmentation function
\begin{equation}
D_{\Lambda_c/c}(z) =  \delta(1-z).
\label{delta}
\end{equation}
The use of the delta fragmentation function implies that the 
$\Lambda_c$ is produced with the same momentum carried by the 
fragmenting $c$-quark. This mechanism for fragmentation has been used 
to simulate the coalescence of the $c$-quark, produced in a hard 
interaction, with light valence quarks coming from the initial 
hadrons \cite{ramona4}.

\begin{figure}[b] 
\psfig{figure=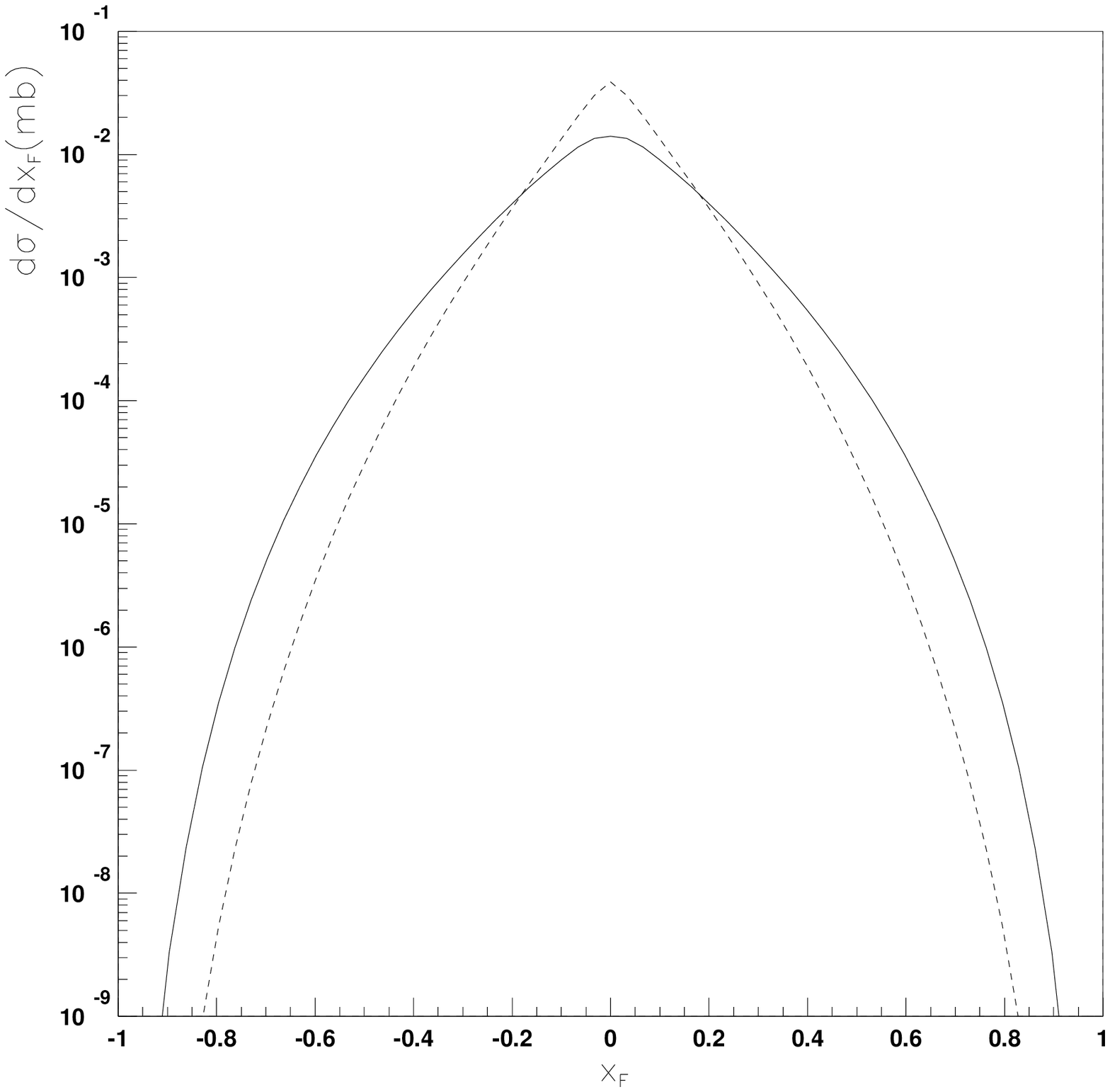,height=6.0in} 
\caption{}
\label{pf-pp} 
\end{figure}

\begin{figure}[b] 
\psfig{figure=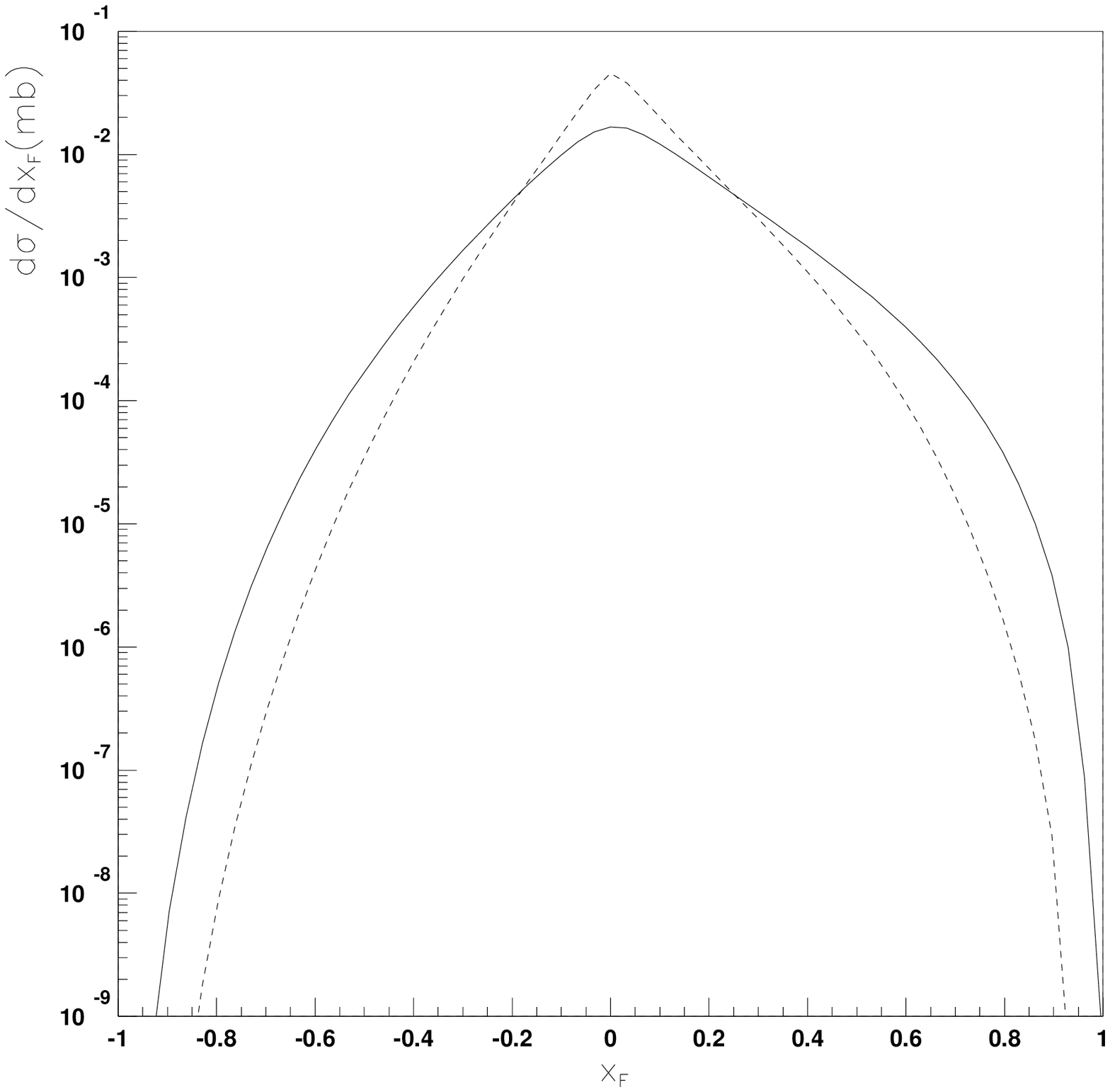,height=6.0in} 
\caption{}
\label{pf-pip} 
\end{figure}

In Figs. \ref{pf-pp} and \ref{pf-pip} we show the parton fusion 
distributions obtained for the two fragmentation functions 
in $pp$ and $\pi^- p$ interactions respectively. Notice that 
the $\Lambda_c$ and the $\overline{\Lambda_c}$ 
parton fusion distributions are equal at LO. However, at NLO 
there is an small $\bar{c}/c$ asymmetry due to the fact that to 
this order in perturbative theory the cross section for the production 
of a quark differs from the cross section for the production of an 
anti-quark \cite{nlo0} and consequently a small 
$\overline{\Lambda}_c/\Lambda_c$ asymmetry is present.

\subsection{$\Lambda _c$ and $\overline{\Lambda}_c$ production 
{\it via} recombination in $pp$ and $\pi^- p$ interactions}
\label{recom}

The method introduced by K.P. Das and R.C. Hwa for meson recombination 
\cite{das-hwa} was extended by J. Ranft \cite{ranft} to describe single 
particle distributions of baryons in $pp$ collisions. In a recent work 
R.C. Hwa \cite{hwa-prd} studied $D^{\pm}$ inclusive production using the 
valon recombination model.

In recombination, all products of the reaction appearing in the forward 
region ($x_F > 0$) are thought as coming from the beam particle 
fragmentation while hadrons produced in the backward region ($x_F < 0$)
come from the target fragmentation. Since the two reactions we are 
considering here have the same target particles, then the 
recombination $\Lambda_c$ and 
$\overline{\Lambda}_c$ inclusive $x_F$ distributions 
are the same for both reactions in the backward region. In the forward 
region, however, they must be different since the outgoing hadron 
is produced from the debris of different initial particles.
Hence it is sufficient to study $\Lambda_c$ and $\overline{\Lambda}_c$ 
production by recombination in the forward region for $pp$ and $\pi^-p$ 
collisions, being the backward $x_F$ distributions symmetrical with respect 
to the forward $x_F$ distribution in $pp$ interacctions in both reactions 
under study.

The invariant $x_F$ inclusive distribution for $\Lambda_c\;\left(
\overline{\Lambda}_c\right)$ production in $pp$ interactions in 
the forward region is given by
\begin{equation} 
\frac{2 E}{\sigma^{rec}\sqrt {s}}\frac{d\sigma^{rec}
}{dx_F}=\int_0^{x_F}\frac{dx_1}{x_1}\frac{dx_2}{x_2}\frac{dx_3
}{x_3}F_3^{\Lambda_c\;\left(\overline{\Lambda}_c\right)}
\left( x_1,x_2,x_3\right) R_3\left( x_1,x_2,x_3,x_F\right)
\label{rec}
\end{equation}
where $E$ is the energy of the produced hadron, $\sqrt{s}$ is the 
center of mass energy, $R_3\left( x_1,x_2,x_3,x_F\right)$ is the 
recombination function and $F_3^{\Lambda_c\;\left(\overline{\Lambda}_c
\right)}\left( x_1,x_2,x_3\right)$ is the 
three quark distribution function. $x_i$, i=1, 2, 3  is the 
momentum fraction of the $i^{th}$ quark with respect to the proton.

Following the approach of Ref. \cite{ranft}, the three quark 
distribution function is assumed to be of the form 
\begin{equation}
F_3^{\Lambda_c\left(\overline{\Lambda}_c\right)} \left( x_1,x_2,x_3 \right) = 
\beta_p F_{q_1}^{p}\left(x_1\right)F_{q_2}^{p}\left(x_2\right)
F_{q_3}^{p}\left(x_3\right)
\left(1-x_1-x_2-x_3\right)^{\gamma_p}
\label{3quark-p}
\end{equation}
In eq. \ref{3quark-p}, $F_{q_i} = x_i q_i(x_i)$ is the single quark 
distribution of the $i^{th}$ valence quark in the produced 
$\Lambda_c \; (\overline{\Lambda}_c)$ inside the proton and the coefficients 
$\beta_p$ and $\gamma_p$ are fixed using the consistency condition
\begin{eqnarray}
F_{q}\left(x_i\right) & = & \int_0^{1-x_i}dx_j \int_0^{1-x_i-x_j}
dx_k \:F_3^{\Lambda_c\left(\overline{\Lambda}_c\right)} 
\left( x_1,x_2,x_3 \right)\:. \nonumber \\
                      &   &i,j,k = 1,2,3 
\label{consist}
\end{eqnarray}
which must be valid for the valence quarks in the proton.

In eq. \ref{3quark-p} we use the single quark GRV-LO 
distributions \cite{grv-p}. Since the GRV-LO distributions 
are functions of the momentum fraction $x$ and the momentum scale 
$Q^2$, then our 
$F_3^{\Lambda_c/\overline{\Lambda}_c} \left( x_1,x_2,x_3 \right)$ 
also depends on $Q^2$. We use $Q^2 = 4m_c^2$ with $m_c = 1.5$ GeV 
as in Ref. \cite{nosotros}. Notice that the scale in the recombination 
is fixed at a different value than in parton fusion. In fact, in parton 
fusion the scale $Q^2$ is fixed at the vertices of the Feynman diagrams 
involved in the perturbative part of the calculation while in recombination 
this parameter must be chosen in such a way that small variations in its 
value does not change appreciably the charmed sea of the initial hadron.

Concerning the single-quark contributions to the three quark 
distribution function, some comments are in order. In $pp$ 
interactions, contributions to the $\Lambda_c$ inclusive $x_F$ 
distribution in the large $x_F$ region come mainly from 
Valence-Valence-Sea (VVS) recombination processes, being other processes 
involving more than one sea-flavor in the recombination completely 
negligible due to the fast fall of the sea-quark distributions. 
Conversely, in the small $x_F$ region, contributions of Valence-Sea-Sea 
(VSS) and Sea-Sea-Sea (SSS) recombination processes are important 
since sea-quark distributions are peaked at $x_F$ about zero. In 
this region, the later processes dominate largely over VVS 
recombination. Hence, in order to have a more acurate prediction for the 
$\Lambda_c/\overline{\Lambda}_c$ asymetry at small $x_F$, VSS and SSS 
processes must be included in the calculation of the $\Lambda_c$ 
cross section. $\overline{\Lambda}_c$ production in $pp$ interactions 
proceeds only through SSS recombination.

For the recombination function we take
\begin{equation}
R_3\left( x_1,x_2,x_3\right) =\alpha \frac{(x_1x_2)^{n_1}x_3^{n_2}}
{x_F^{n_1+n_2-1}}
\delta \left(x_1+x_2+x_3-x_F\right)   
\label{eq7}
\end{equation}
allowing in this way for a different weight 
for the heavy $c$ ($\bar{c}$) quark (marked with index 3) 
than for the light $u$ ($\bar{u}$) and $d$ ($\bar{d}$) quarks 
(indexed 1 and 2 respectively), as 
suggested in Ref. \cite{hwa-prd} in conection with charmed meson 
production in the valon model.

The constant $\alpha$ in eq. \ref{eq7} is fixed by the condition 
\cite{nosotros} 
\begin{equation}
\frac{1}{\sigma^{rec}} 
\int_0^1 dx_F \frac{d\sigma^{rec}}{dx_F} = 1
\label{eq7b}
\end{equation}
for $\Lambda_c$ inclusive production. Using the same value for 
the parameter $\alpha$ obtained in eq. \ref{eq7b} 
to normalize the $\overline{\Lambda}_c$ $x_F$ distribution, the anti-particle 
cross section is given relative to the particle cross section. In this way 
$\sigma^{rec}_{\Lambda_c}$ has the physical interpretation of the 
$\Lambda_c$ total recombination cross section and it should be fixed 
by experimental data.

\begin{figure}[b] 
\psfig{figure=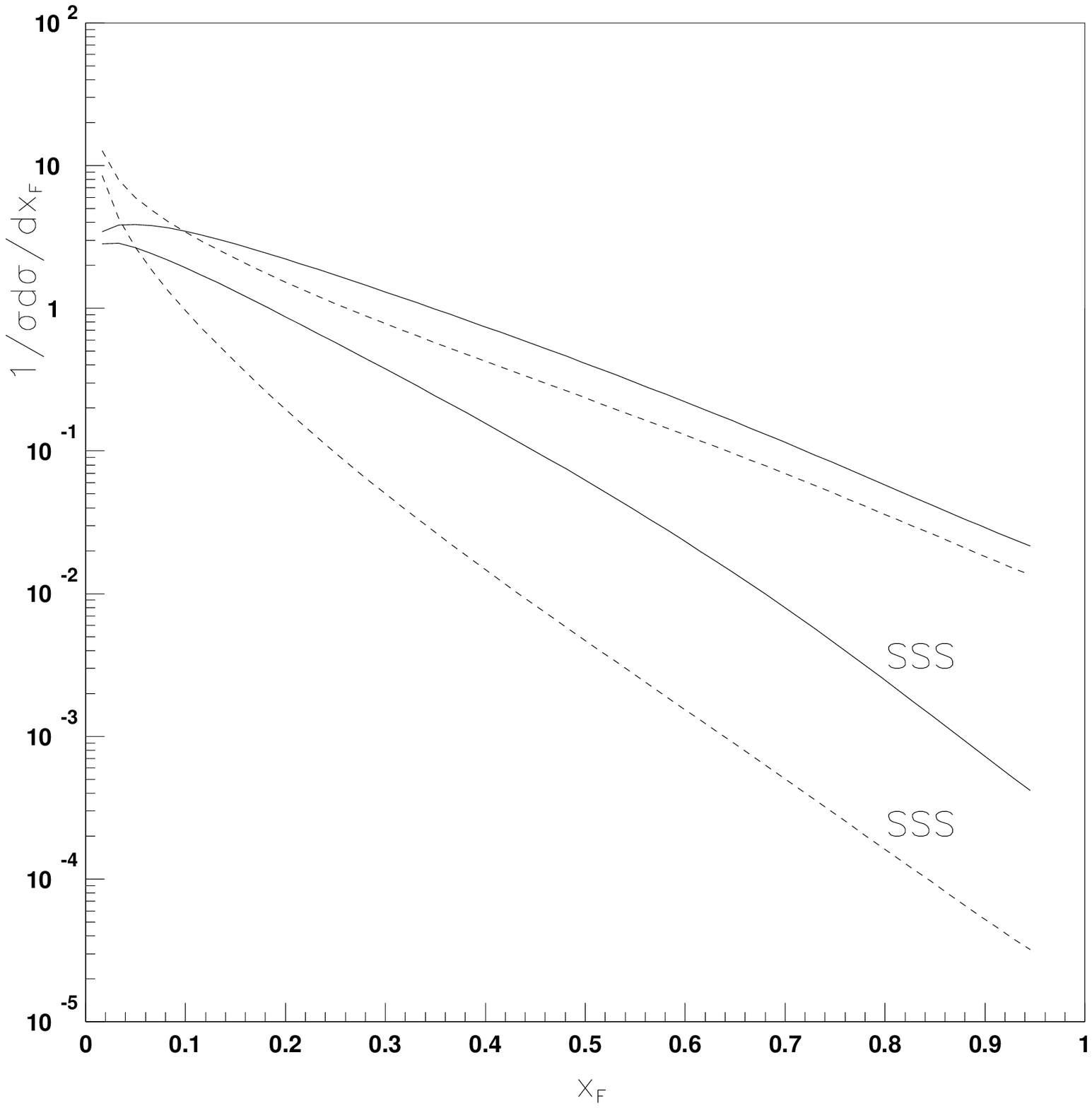,height=6.0in} 
\caption{}
\label{recomb} 
\end{figure}

In Fig. \ref{recomb} we show the $\Lambda_c$ and $\overline{\Lambda}_c$ 
$x_F$ inclusive distribution for $n_1 = n_2= 1$ \cite{ranft} 
and $n_1 = 1$, $n_2 = 5$ \cite{hwa-prd}
with $\beta_p = 75$ and $\gamma_p = -0.1$ as given in 
Ref. \cite{nosotros}.

In $\pi^-p$ interactions in the beam fragmentation region both $\Lambda_c$ 
and $\overline{\Lambda}_c$ inclusive $x_F$ distributions are given by 
formulas formally identical 
to that of eq. \ref{rec}. As the $\pi^-$ has an $\bar{u}$ and a $d$ 
valence quarks, the recombination processes involved in $\Lambda_c$ 
as well as in $\overline{\Lambda}_c$ production are VSS and SSS. On the 
other hand, since the GRV-LO \cite{grv-pi} valence distributions and 
quark and anti-quark sea distributions in the pion are equal, then the 
$\Lambda_c$ and $\overline{\Lambda}_c$ inclusive $x_F$ distributions are 
the same. Notice that $\Lambda_c$ production in $\pi^-p$ interactions 
in the proton fragmentation region is double leading and $\overline
{\Lambda}_c$ is non-leading whereas in the beam fragmentation region 
both particle and anti-particle production are leading, so in this case 
a $\Lambda_c/\overline{\Lambda}_c$ asymmetry is expected for $x_F < 0$.

\section{$\Lambda_c/\overline{\Lambda}_c$ production asymmetry in the 
recombination two component model}

The total inclusive $x_F$ distributions are obtained by adding the 
parton fusion and recombination contributions given by eqs. \ref{sig-qcd} 
and \ref{rec} respectively. Its general form is
\begin{equation}
\frac{d\sigma^{tot}}{dx_F}\left|_{\Lambda_c(\overline{\Lambda}_c)} \right.= 
\frac{d\sigma^{pf}}{dx_F} + 
\sigma^{rec}_{\Lambda_c}
\frac{d\sigma^{rec}}{dx_F}\left|_{\Lambda_c(\overline{\Lambda}_c)}
\right.
\label{lam-pp}
\end{equation}
where we have uncovered the parameter $\sigma^{rec}_{\Lambda_c}$ 
hidden in the $d\sigma^{rec}/dx_F$ definition of eq. \ref{rec}. 
Let us stress that $\sigma^{rec}_{\Lambda_c}$ is the only free parameter 
in the model.

The $\Lambda_c/\overline{\Lambda}_c$ production asymmetry is obtained 
by replacing the corresponding total inclusive distribution given 
by eq. \ref{lam-pp} in eq. \ref{eq0}. In this way we obtain 
\begin{equation}
A_{\Lambda_c/\overline{\Lambda}_c} \left( x_F \right) = 
\sigma^{rec}_{\Lambda_c} \frac{d\sigma^{rec}/dx_F \mid_{\Lambda_c} - 
d\sigma^{rec}/dx_F \mid_{\overline{\Lambda}_c}}
{2 d\sigma^{pf}/dx_F + \sigma^{rec}_{\Lambda_c}
\left[ d\sigma^{rec}/dx_F \mid_{\Lambda_c} +
d\sigma^{rec}/dx_F \mid_{\overline{\Lambda}_c} \right]}
\label{asym-pp}
\end{equation}
for the asymmetry as a function of $x_F$.

It should be noted that pQCD at NLO predicts a bigger $\overline{\Lambda}_c$ 
than $\Lambda_c$ production at large values of $x_F$. Since the asymmetry 
coming from pQCD is of the order of 10 $\%$ at $x_F \approx 1$ and 
in the large $x_F$ region the recombination component is several orders 
of magnitude bigger than the perturbative part, this contribution to 
the asymmetry is completely negligible.

In $\pi^-p$ interactions in the forward region ($x_F > 0$), the 
$\Lambda_c$ and the $\overline{\Lambda}_c$ recombination cross sections 
are equal, then our model predicts zero or a small negative asymmetry due 
to NLO pQCD effects. This prediction is consistent with the 
ratio $N(\Lambda_c)/N(\overline{\Lambda}_c) = 
0.99 \pm 0.16$ measured by the ACCMOR collaboration \cite{accmor} which 
indicates no asymmetry ($A = -0.002 \pm 0.16$). Also a preliminary 
analysis in 500 GeV/c $\pi^-$-nucleus interactions from the E791 
collaboration \cite{e791} is consistent with zero asymmetry in the 
region $0 \leq x_F \leq 0.55$.

In the backward region, the same E791 preliminary analysis shows a 
large asymmetry, reaching a value of $A \approx 0.4$ at $x_F \approx -0.13$ 
but with large error bars \cite{e791}. The ratio 
$N(\Lambda_c)/N(\overline{\Lambda}_c) = 1.17 \pm 0.08$ 
($A = 0.078 \pm 0.034$) measured by the E791 
collaboration \cite{e791} is described by our model with a 
recombination cross section $\sigma^{rec}_{\Lambda_c} \approx 1.1 
\sigma^{pf}$ and $n_1 = 1$, $n_2 = 5$ in the recombination function  
of eq. \ref{eq7} in the Peterson fragmentation scheme. Here 
$\sigma^{pf} = \int_0^1{dx_F d\sigma^{pf}/dx_F}$. Both Peterson 
and Delta fragmentation in combination with $n_1=n_2=1$ in the 
recombination function tend to give a more slowly growing asymmetry 
with $x_F$.

The E769 experiment \cite{e769} found a lower limit of 
$A = 0.6$ in the forward region with 
a 250 GeV/c incident $\pi^-$, $K$ and $p$ beam on a multifold 
target of Be, Cu, Al and W. However, altough this experiment presents 
an evidence of $\Lambda_c/\overline{\Lambda}_c$ asymmetry, the value 
quoted in Ref. \cite{e769} for the asymmetry is not significative 
due to its very low statistic.

\begin{figure}[b] 
\psfig{figure=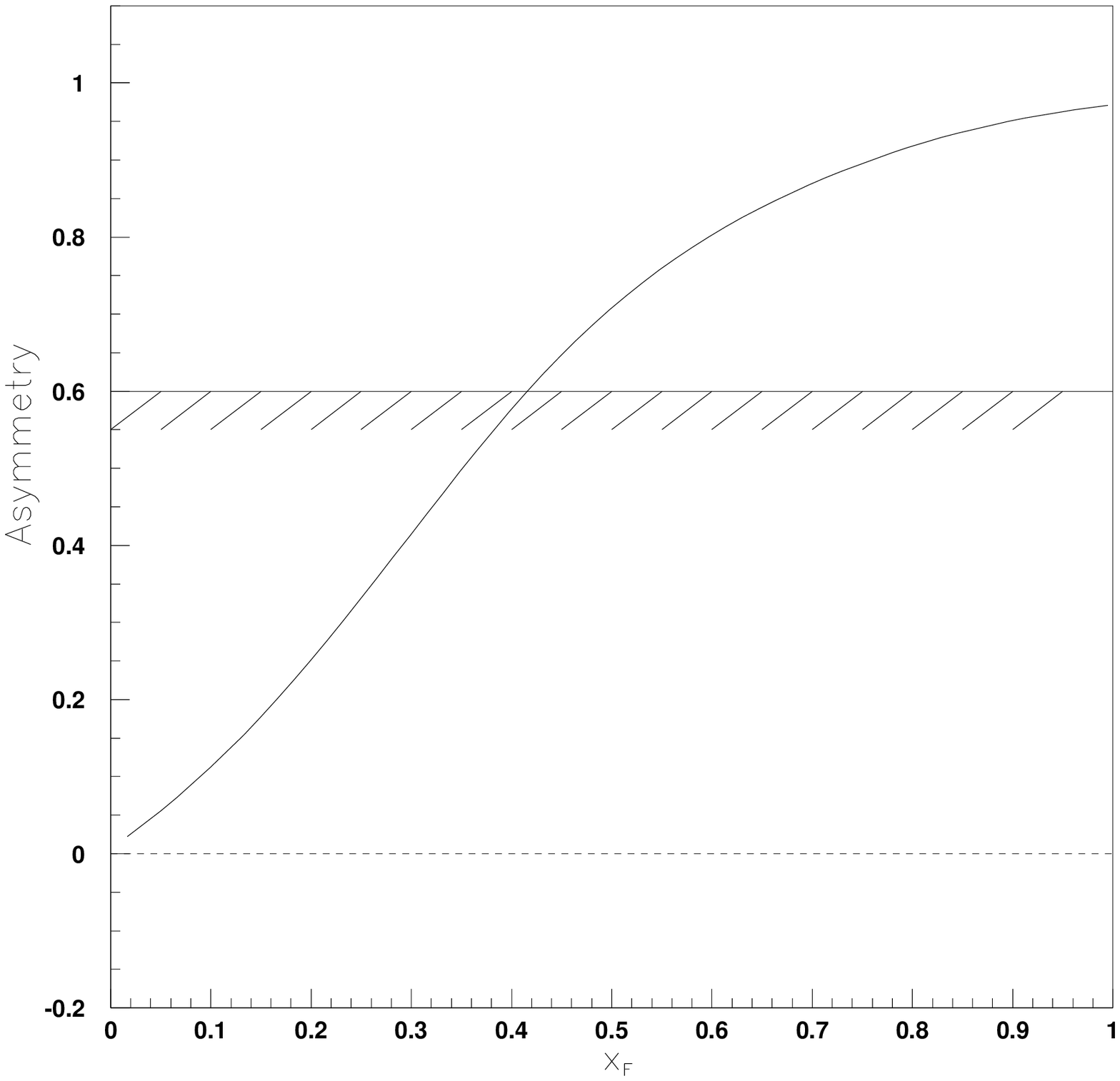,height=6.0in}
\caption{} 
\label{asym} 
\end{figure}

\begin{figure}[b] 
\psfig{figure=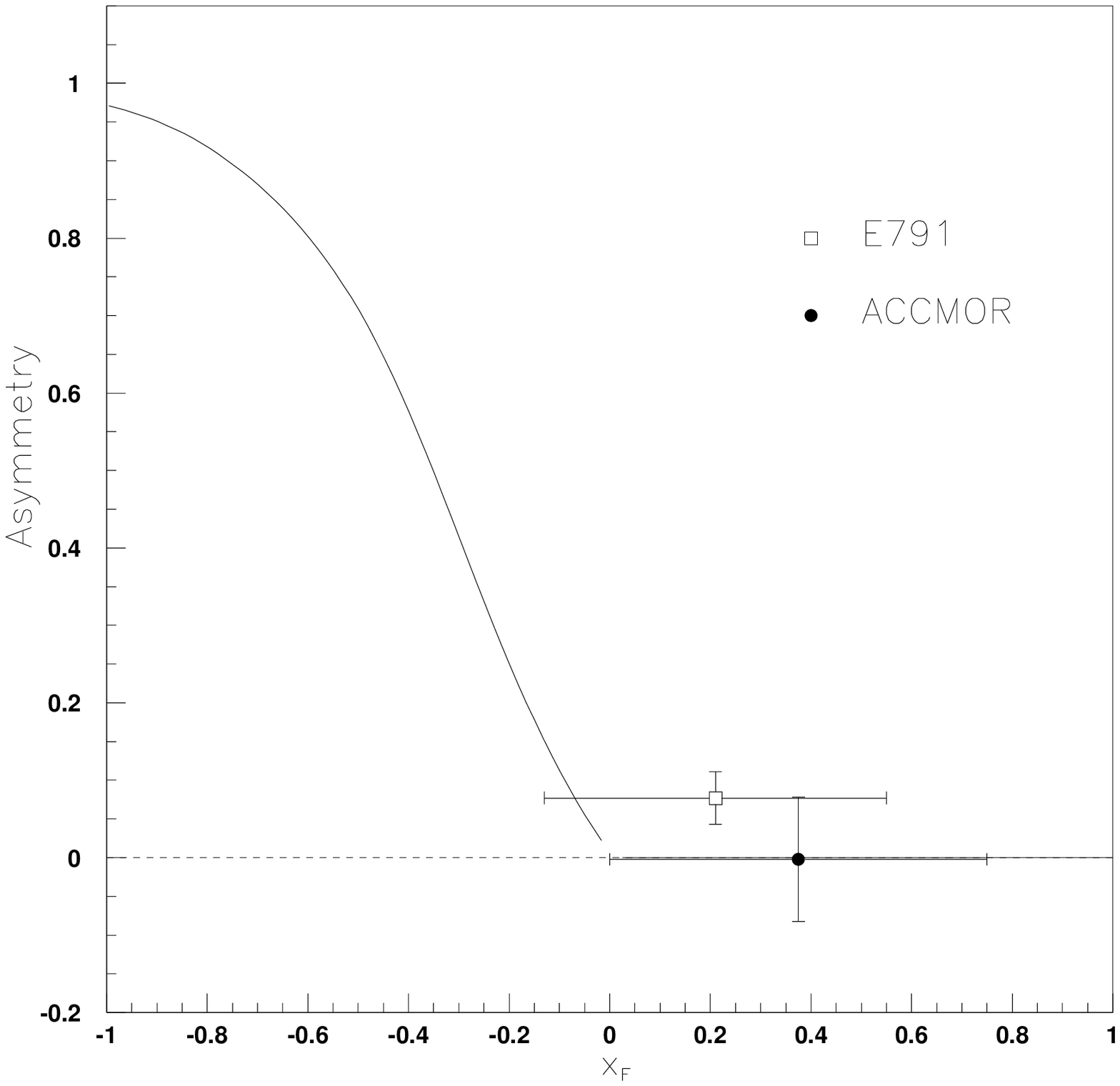,height=6.0in}
\caption{} 
\label{asym2} 
\end{figure}

In Figs. \ref{asym} and \ref{asym2} we show the asymmetry as a function 
of $x_F$ predicted by the recombination two component model in $pp$ and 
$\pi^- p$ interactions respectively with the recombination normalization 
suggested by the preliminary analisys from the E791 collaboration.

\section{Conclusions}

In this work we have presented the predictions of the recombination 
two component model for the $\Lambda_c/\overline{\Lambda}_c$ 
production asymmetry in $pp$ and $\pi^-p$ interactions.

The model seems to describe adequately the scarce data available on 
the subject but, of course, more experimental data are nedeed not 
only on $\Lambda_c/\overline{\Lambda}_c$ asymmetries as well as 
on $\Lambda_c$ and $\overline{\Lambda}_c$ $x_F$ distributions 
in order to do a meaningful comparison.

Using the preliminary analysis of the E791 collaboration \cite{e791}, 
the model predicts that the recombination cross section is as big as 
the parton fusion cross section, giving a clear idea about the 
extent to which non-perturbative 
contributions are important in charm hadroproduction.

It should be noted that if the production asymmetry is independent 
of the energy, as indicated by measurements of charmed meson/
anti-meson asymmetries, the model predicts that the ratio of 
the recombination to the parton fusion cross section does not depend 
on the energy and could bring about important information about the 
non-perturbative processes in heavy hadron production.

In the recombination two component model, the part responsible 
for the asymmetry is recombination. It involves different kinds of 
processes for $\Lambda_c$ than for 
$\overline{\Lambda}_c$ production, giving $x_F$ inclusive distributions 
which are different in shape for the particle than for the anti-particle.
 
Notice that strong diquark effects are present in the model for 
$\Lambda_c$ production in the proton fragmentation region in the two 
reactions considered in this paper. In fact, since the $u$ and $d$ are 
valence quarks in the proton, the $ud$ diquark carries a large 
amount of the proton momentum when it is liberated in the collision. This 
large momentum is then transferred to the $\Lambda_c$ when the $ud$ 
diquark recombines with a sea $c$-quark to form the outgoing particle.

Another model which has been used to make predictions on the 
$\Lambda_c/\overline{\Lambda}_c$ production asymmetry is the intrinsic 
charm two component model \cite{ramona4}. In this model the enhancement 
in the $\Lambda_c$ over the $\overline{\Lambda}_c$ cross section 
is due to the coalescence of $u$, $d$ and $c$ quarks coming from the 
$\left|uudc\bar{c}\right>$ Fock state of the proton.

However there are important differences between predictions obtained 
with one or another model.

In fact, one of the possibly most interesting features which distinguish 
the recombination from the intrinsic charm two component model is 
the ability of the first to produce large asymetries in the small $x_F$ 
region. This diference between the two models is due to the fact 
that recombination of sea and valence quark produces a $\Lambda_c$ 
$x_F$ distribution peaked close to zero whereas the intrinsic charm 
model gives a $\Lambda_c$ $x_F$ distribution slowly growing from 
zero at $x_F = 0$ to its maximum at $x_F$ about $0.6$.

Finally, it should be noted that the two models seem to describe 
adequately the shape of the $\Lambda_c$ inclusive $x_F$ distribution 
(see Refs. \cite{ramona4,nosotros}) but both models predict very 
different forms for the $\Lambda_c/\overline{\Lambda}_c$ asymmetry, 
so a meaningful comparison between models and experimental data 
should be done with data on asymmetries as well as on $x_F$ 
distributions.


\section*{Acknowledgments} 


We gratefully acknowledge very useful discussions with J. Appel. 
We wish to thank also Centro Brasileiro de Pesquisas F\'{\i}sicas 
(CBPF) for the hospitality extended to us.
This work was partially supported by the Centro Latino Americano de 
F\'{\i}sica (CLAF). J.M. is supported by Funda\c{c}\~ao de Amparo 
\`a Pesquisa do Estado de Rio de Janeiro (FAPERJ).



%
\newpage

\section*{Figure Captions}
 
\begin{itemize}
\item
[Fig. 1:] Feynman diagrams involved in the LO calculation of the 
parton fusion cross section.
\item
[Fig. 2:] Parton fusion cross section for $pp$ at 500 GeV/c energy 
beam. Peterson (dashed line) and Delta (full line) fragmentation 
are shown.
\item
[Fig. 3:] Parton fusion cross section for $\pi^-p$ at 500 GeV/c energy 
beam. Peterson (dashed line) and Delta (full line) fragmentation 
are shown.
\item
[Fig. 4:] Recombination cross section in $pp$ collisions for 
$n_1=1$, $n_2=5$ (full line) and $n_1=n_2=1$ (dashed line). 
Curves marked "SSS" are the corresponding $\overline{\Lambda}_c$ 
distributions. The $\Lambda_c$ distribution is normalized to 
unity and the $\overline{\Lambda}_c$ 
distribution is normalized relative to the first one.
\item 
[Fig. 5:] $\Lambda_c/\overline{\Lambda}_c$ asymmetries in $pp$  
interactions for Peterson fragmentation and $n_1 = 1$, $n_2 = 5$ in the recombination function calculated for the recombination cross section 
suggested by the E791 preliminary data. The lower limit obtained by 
the E769 (Ref. [4]) experiment is shown.
\item 
[Fig. 6:] Same as in Fig. 1 for $\pi^- p$ interactions. Experimental 
points are calculated from the $N(\Lambda_c)/N(\overline{\Lambda}_c)$ 
ratios measured in Refs. [5] and [19]
\end{itemize}
\newpage


\begin{thebibliography}{99}

\bibitem{ramona0} T. Carter, in proceedings of DPF'94, the 8$^{th}$ 
meeting of the APS Division of Particles and Fields, Albunquerque, NM, 
1994, Seidel Ed., M. Aguilar Benitez {\it et al.}, Phys Lett {\bf B 161}, 
400 (1985) and Z. Phys. {\bf C 31}, 491 (1986), S. Barlag {\it et al.}, 
Z. Phys. {\bf C 49}, 555 (1991), M. Adamovich {\it et al.}, 
Phys Lett {\bf B 305}, 402 (1993) and G.A. Alves {\it et al.}, Phys. Rev. 
Lett. {\bf 72}, 812 (1994).

\bibitem{e791dd} E791 collaboration (E. Aitala {\it et al.}), 
Phys. Lett. {\bf B 371}, 157 (1996).

\bibitem{ramona1} P. Chauvat {\it et al.} Phys. Lett. {\bf B 199}, 
304 (1987),  G. Bari {\it et al.}, Nuovo Cimento, {\bf 104 A}, 57 (1991).


\bibitem{e769} E769 Collaboration (G. Alves {\it et al.}), 
Phys. Rev. Lett. {\bf 77}, 2388 (1996).

\bibitem{e791} A.M. Halling, Il Nuovo Cim. {\bf 109 A}, 617 (1996) and 
S. Kwan (E791 Collaboration), private communication.

\bibitem{ramona2} G. Bari {\it et al.}, Nuovo Cimento {\bf 104 A}, 
787 (1991).

\bibitem{ramona3} S.F. Biagi {\it et al.}, Z. Phys. {\bf C 28}, 175 (1985); 
R. Werding, WA89 Collaboration, in proceedings of ICHEP94, Glasgow.

\bibitem{nlo0} P. Nason, S. Dawson and R.K. Ellis, Nucl. Phys. 
{\bf B 327}, 49 (1989).

\bibitem{hwa-prd} R.C. Hwa, Phys. Rev. {\bf D 51}, 85 (1995).

\bibitem{ramona4}R. Vogt and S.J. Brodsky, Nucl. Phys. {\bf B 478}, 311 (1996).

\bibitem{nosotros} J.C. Anjos, G. Herrera, J. Magnin and F.R.A. 
Sim\~ao, Phys. Rev. {\bf D 56}, 394 (1997).

\bibitem{nlo} R. Vogt, Z. Phys. {\bf C 71}, 475 (1996).

\bibitem{vbh-npb} J. Babcock, D. Sivers and S. Wolfram, Phys. Rev. {\bf D 
18}, 162 (1978), B.L. Combridge, Nucl. Phys. {\bf B 151}, 429 (1979), 
R.K. Ellis, Fermilab-Conf-89/168-T (1989), I. Inchiliffe, Lectures at 
the 1989 SLAC Summer Institute, LBL-28468 (1989).

\bibitem{grv-p} M. Gl\"{u}ck, E. Reya and A. Vogt, Z. Phys. {\bf C 53}, 
127 (1992).

\bibitem{grv-pi} M. Gl\"{u}ck, E. Reya and A. Vogt, Z. Phys. {\bf C 53}, 
651 (1992).

\bibitem{peterson} C. Peterson, D. Schlatter, J. Schmitt and P. Zerwas, 
Phys. Rev. {\bf D 27}, 105 (1983).

\bibitem{das-hwa} K.P. Das and R.C. Hwa, Phys. Lett. {\bf B 68}, 
459 (1977).

\bibitem{ranft} J. Ranft, Phys. Rev. {\bf D 18}, 1491 (1978).

\bibitem{accmor} ACCMOR Collaboration (S. Barlag {\it et al.}), 
Phys. Lett. {\bf B 247}, 113 (1990).

\end{thebibliography}
\end{document}